\documentclass[twocolumn,superscriptaddress]{revtex4-1}
\usepackage{color}
\usepackage{graphicx}
\usepackage{epstopdf}
\usepackage{amsmath}
\usepackage{amssymb}
\usepackage[colorlinks,linkcolor=blue,anchorcolor=blue,citecolor=blue,urlcolor=blue]{hyperref}

\newcommand{\bea}{\begin{eqnarray}}
\newcommand{\eea}{\end{eqnarray}}

\newcommand{\ie}{\textit{i.e.{ }}}
\newcommand{\eg}{\textit{e.g.{ }}}

\newcommand{\etal}{\textit{et al.{ }}}
\newcommand{\me}{\mathrm{e}}

\begin{document}

\title{Entanglement in a second order topological insulator on a square lattice}
\author{Qiang Wang}
\affiliation{National Laboratory of Solid State Microstructures $\&$ School of Physics, Nanjing University, Nanjing, 210093, China}
\author{Da Wang} 
\affiliation{National Laboratory of Solid State Microstructures $\&$ School of Physics, Nanjing University, Nanjing, 210093, China}
\author{Qiang-Hua Wang} 
\affiliation{National Laboratory of Solid State Microstructures $\&$ School of Physics, Nanjing University, Nanjing, 210093, China}
\affiliation{Collaborative Innovation Center of Advanced Microstructures, Nanjing University, Nanjing 210093, China}

\begin{abstract}
  In a $d$-dimensional topological insulator of order $d$, there are zero energy states on its corners which have close relationship with its entanglement behaviors. We studied the bipartite entanglement spectra for different subsystem shapes and found that only when the entanglement boundary has corners matching the lattice, exact zero modes exist in the entanglement spectrum corresponding to the zero energy states caused by the same physical corners. We then considered finite size systems in which case these corner states are coupled together by long range hybridizations to form a multipartite entangled state. We proposed a scheme to calculate the quadripartite entanglement entropy on the square lattice, which is well described by a four-sites toy model and thus provides another way to identify the higher order topological insulators from the multipartite entanglement point of view.
\end{abstract}
\maketitle
\section{introduction}
Entanglement and topological states have close relationship, which is extensively studied in recent years. \cite{Kitaev_PRL_2006, Levin_PRL_2006, Ryu_PRB_2006, Li_PRL_2008, Fidkowski_PRL_2010, Pollmann_PRB_2010, Yao_PRL_2010, Turner_PRB_2011, Zhang_PRL_2011, Huang_PRB_2012, Qi_PRL_2012, Jiang_PRL_2013, Chandran_PRL_2014, Wang_PRB_2015, Zeng_a_2015, Laflorencie_PR_2016, Koch-Janusz_PRB_2017} Generally speaking, a nontrivial structure of the entanglement spectrum, \eg degeneracy of the many body one or zero modes of the single particle one, always indicates a nontrivial topological state, and vice versa. \cite{Ryu_PRB_2006, Li_PRL_2008, Fidkowski_PRL_2010, Qi_PRL_2012} The key point is that the entanglement boundary in some sense mimics the physical boundary which is further related to the topological property via the usual bulk-boundary correspondence: a $d$-dimensional topological insulator has gapless states on its $(d-1)$-dimensional boundaries. \cite{Hatsugai_PRL_1993, Qi_PRB_2006, Qi_PRB_2008, Schnyder_PRB_2008}

Very recently, a generalization of the topological insulator called higher order topological insulator is theoretically predicted and observed in experiments. \cite{Benalcazar_S_2017, Benalcazar_PRB_2017, Schindler_SA_2018, Song_PRL_2017, Parameswaran_P_2017, Xu_a_2017, Ezawa_PRL_2018, Ezawa_PRB_2018, Kunst_PRB_2018, Schindler_a_2018} Different from the conventional topological insulators, the gapless boundary states now exist on the $(d-n)$-dimensional boundaries for a $n$-th order topological insulator. The widely used entanglement spectrum with cylindrical (smooth) bipartite scheme can not be directly applied to identify such higher order topological insulators due to the missing of $(d-1)$-dimensional gapless boundary states. However, a straightforward generalization can be conceived by choosing the subsystem with $(d-n)$-dimensional (non-smooth) entanglement boundaries, in which case an additional contribution from the non-smooth corners \cite{Laflorencie_PR_2016} would lead to a nontrivial entanglement spectrum and thus identify the higher order topological insulators. Such a scheme is proposed in a similar way called nested entanglement spectra by Schindler \etal in Ref.~\onlinecite{Schindler_SA_2018}. In this work, we focus on a special case with $n=d$ on the square lattice to check the validity of this conjecture and further examine the effect of different subsystem shapes on the entanglement spectra. It turns out that only when the corners of the subsystem match the full lattice, exact zero modes exist in the entanglement spectrum which correspond to the zero energy corner states caused by physical boundaries.

On the other hand, in a real topological system with finite size, its zero energy boundary states can be coupled together by long range hybridizations and thus also contribute to the total entanglement spectrum/entropy, which was proposed to detect the topological degeneracy induced by boundaries in conventional (first order) topological insulators. \cite{Wang_PRB_2015} Here, for the $d$-th order topological insulator with open boundary condition, the zero energy corner states are hybridized to form a fully multipartite entangled state. Focusing on the square lattice case, we propose a scheme to calculate the quadripartite entanglement entropy as a subleading correction of the area law. The obtained result shows a universal value which is well described by a toy model with only four sites and thus directly identifies the existence of these zero energy corner states. 

\section{model and bipartite entanglement spectra}

\begin{figure}
    \includegraphics[width=0.5\textwidth]{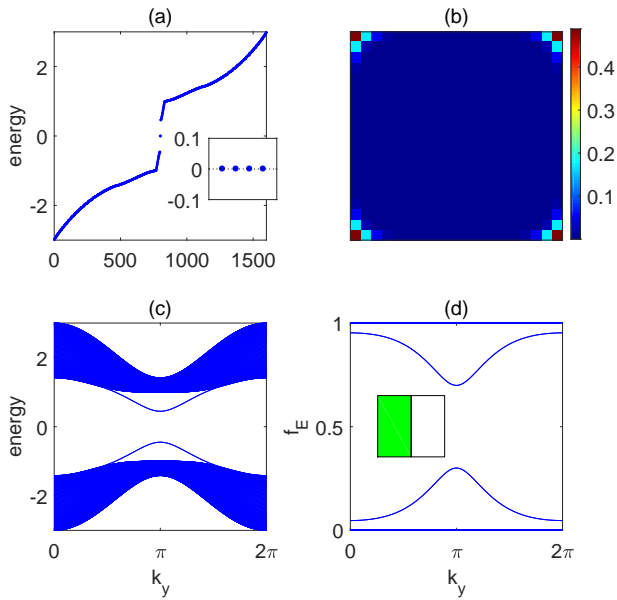}
    \caption{\label{fig:model} (a) Energy spectrum of a $20\times20$ lattice with open boundary condition. The inset shows the enlarged plot near the zero energy. (b) Local density of states of these zero energy modes. (c) Energy spectrum on a cylinder with the y direction periodic. (d) Entanglement occupation $f_E$ of the cylindrical bipartite scheme as shown in the inset where green region denotes the subsystem. }
\end{figure}

We adopt the model from Ref.~\onlinecite{Schindler_SA_2018} defined on the square lattice,
\begin{eqnarray} \label{eq:model}
  H &=& \sum_{i}\sum_{e=x,y} c_i^\dag\left( t\sigma_0\tau_z + i\lambda \sigma_e\tau_x + \Delta d_e \sigma_0\tau_y \right)c_{i+e} + h.c. \nonumber\\ 
    &+& \sum_i c_i^\dag m\sigma_0 \tau_z c_i
\end{eqnarray}
where $c_i$ is a four component (two spins and two orbits) fermionic annihilation operator with $i$ the lattice site index, $\sigma_{x,y,z}$/$\tau_{x,y,z}$ are Pauli matrices acting in spin/orbital space while $\sigma_0$ is the rank two identity matrix. In the last term, $d_e=1,-1$ for $e=x,y$, respectively. When $\Delta=0$, this model respects a time reversal symmetry represented by $\mathcal{T}=\sigma_y\tau_0\mathcal{K}$ ($\mathcal{K}$ means complex conjugate) and goes back to the usual topological insulator with gapless edge states. \cite{Qi_PRB_2008} When $\Delta\ne0$, the time reversal symmetry is broken and thus gaps out the edge states. But at each corner matching the square lattice, one zero energy mode is left as a result of the nonzero bulk quadrupole moment. \cite{Benalcazar_S_2017} Such a picture is seen in Fig.~\ref{fig:model}(a) by solving the energy spectrum in a $20\times20$ lattice with open boundary condition. There are four zero modes as shown in the insets, the charge distribution of which are shown in Fig.~\ref{fig:model}(b). Throughout this work, we use the model parameters $m=2t=2\lambda=1$ and $\Delta=0.25$. 

In this work, we are interested in the entanglement behavior of the model Eq.~\ref{eq:model} partly because the entanglement spectrum and entropy always give us valuable information about the topological properties. In practice, we first solve the correlation matrix $\mathcal{C}_{IJ}=\langle c_I^\dag c_J \rangle$ where $I$ and $J$ are grouped indices containing both site and internal degrees. Then, by choosing $I_A\in A$ and $J_A\in A$ where $A$ is a subsystem of the whole lattice, we get a constraint correlation matrix $\mathcal{C}_A$. \cite{Peschel_JPAMG_2003, Ryu_PRB_2006} Its eigenvalues $f_{E,n}$ are called entanglement occupation and directly related to the single particle entanglement spectrum $\xi_n=\ln\left(f_{E,n}^{-1}-1\right)$ which are defined as the energy levels of the reduced Hamiltonian $\mathcal{H}_A$ as appearing in the reduced density matrix $\rho_A=\frac{1}{Z_A}\me^{-\mathcal{H}_A}$. From the entanglement occupation, the entanglement entropy is directly given by $S_E=\sum_n -f_{E,n}\ln f_{E_n} - (1-f_{E,n})\ln(1-f_{E,n}) $. 

\begin{figure}
    \includegraphics[width=0.5\textwidth]{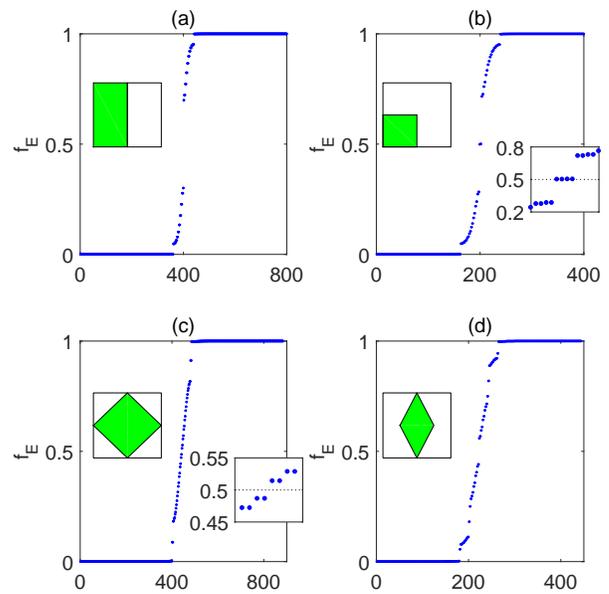}
    \caption{\label{fig:subshape} Entanglement occupation $f_E$ of different bipartite schemes specified by their insets with green regions denoting the subsystems. For clarity, the spectra near $f_E=0.5$ are also enlarged in (b) and (c). The lattice sizes are $20\times20$ in (a,b) and $21\times21$ in (c,d).}
\end{figure}

We first investigate the bipartite entanglement spectra. In topological insulators, the bipartite entanglement spectra are proven to have exact correspondence to the edge states. \cite{Fidkowski_PRL_2010} In special, for a given entanglement boundary, a pair of entanglement occupations $f_{E}=0.5$ correspond to a pair of zero energy states caused by the same physical boundary. A widely used bipartite scheme for the usual topological insulators employs the cylindrical geometry as shown in the inset of Fig.~\ref{fig:model}(d), for which the entanglement spectrum is gapped out similar to the physical edge states in Fig.~\ref{fig:model}(c). Hence the cylindrical bipartite scheme only indicates that the system is not a (first-order) topological insulator. Then a natural question is whether there is a way to identify the second order topological feature? The answer is yes. In fact, from previous studies in conformal field theory, it is known that the non-smooth entanglement boundary would cause a subleading correction to the area law known as corner contributions. \cite{Laflorencie_PR_2016} Here, By noticing that the zero energy states now do not appear on edges but only on corners and following Fidkowski's correspondence, it is thus expected that a non-smooth bipartite scheme containing the same corners in the entanglement boundary would cause half-occupied modes with $f_E=0.5$ in the entanglement spectrum.

In Fig.~\ref{fig:subshape}, we investigate four kinds of subsystem geometries as shown in each inset, including the cylindrical bipartite scheme in (a), square subsystems in (b,c) and a diamond shaped subsystem in (d). In (a) and (d), the entanglement spectra are gapped out indicating the loss of zero energy modes in the two kinds of physical boundaries. While in (b) when the subsystem has four corners exactly matching the square lattice, we obtain four half-occupied modes with $f_{E}=0.5$ corresponding to the four localized zero energy corner states. The case in (c) is more tricky. The entanglement spectrum does show a nontrivial gapless behavior in thermodynamic limit indicating the existence of four branches of extended gapless states not confined to the corners but to the edges, which however does not imply the (first order) topological insulator since the boundary is only specifically-chosen. 

\section{quadripartite entanglement entropy}

\begin{figure}
    \includegraphics[width=0.5\textwidth]{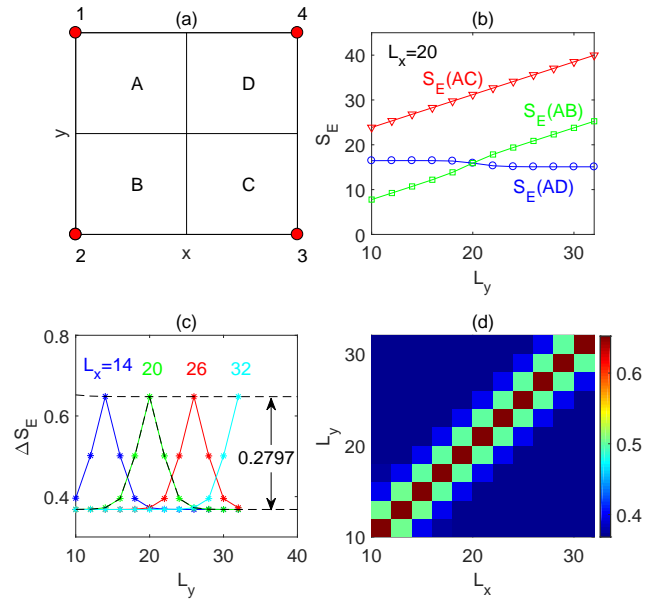}
    \caption{\label{fig:quadripartite} Entanglement entropy in the case of open boundary condition. We divide the whole lattice into four parts labeled by A, B, C and D, respectively as shown in (a). Four red points denote the four corner states. In (b), we plot $S_E(AB)$, $S_E(AD)$ and $S_E(AC)$ as functions of $L_y$ by fixing $L_x=20$. (c) shows $\Delta S_E=S_E(AB)+S_E(AD)-S_E(AC)$ as a function of $L_y$ for different $L_x$. Two dashed lines indicate two limits with $L_x=L_y$ and $L_x\ll L_y$(or $L_x\gg L_y$), respectively. In the large lattice size limit, a universal value $0.2797$ is obtained between the above two limits. In (d) a thorough color plot of $\Delta S_E$ versus $L_x$ and $L_y$ is given.}
\end{figure}

In the above studies, we have assumed periodic boundary condition to calculate entanglement spectra without involving physical boundaries. However, for a real physical system with finite size, \footnote{Exactly speaking, all physical systems are finite and the edge or corner states will be entangled together by long range hybridizations. However, on the other hand, these physical systems are also affected by some environment, which may cause quantum decoherence and thus break down these long range entangled states. Therefore, here, when we say a finite size system, we just mean that the size is smaller than the decoherence length.} the corner states will be hybridized together to form a multipartite entangled state. In the following, we would like to investigate the quadripartite entanglement entropy among the four corner states in our model. Following the idea of obtaining the topological entanglement entropy, \cite{Kitaev_PRL_2006, Levin_PRL_2006} we propose to divide the whole system into four parts denoted by A, B, C and D as sketched in Fig.~\ref{fig:quadripartite}(a). After taking AB, AD and AC as the subsystems to obtain three bipartite entanglement entropies, respectively, we define
\begin{eqnarray}
  \Delta S_E=S_E(AB)+S_E(AD)-S_E(AC)
\end{eqnarray}
which gives a subleading correction to the area law. Physically speaking, $\Delta S_E$ contains two kinds of contributions: one is from the non-smooth corner contributions from the bulk states and the other from the long range quadripartite entanglement among the four corner states.

By numerics we obtain $S_E(AB)$, $S_E(AD)$ and $S_E(AC)$ as functions of $L_y$ by fixing $L_x=20$ as shown in Fig.~\ref{fig:quadripartite}(b). For large enough $L_y$, $S_E(AB)$ and $S_E(AC)$ increase linearly with respect to $L_y$ since the entanglement boundary lengths are proportional to $L_y$. While $S_E(AD)$ converges because the entanglement boundary length only depends on $L_x$. From the three kinds of bipartite entanglement entropies, we obtain $\Delta S_E$ as plotted in Fig.~\ref{fig:quadripartite}(c). For comparison, we also plot curves with different $L_x=14$, $26$ and $32$. More thorough results are shown within a color plot in Fig.~\ref{fig:quadripartite}(d). For all cases with a fixed $L_x$, the curve shows a peak at $L_y=L_x$ and converges when $L_y\gg L_x$ because in which case the four corner states only form a direct product state between $|14\rangle$ and $|23\rangle$ and thus make no contribution to $\Delta S_E$. While the corner contributions are from the short range entangled (gapped) bulk states near the non-smooth entanglement boundaries. Therefore, as long as $L_y$ is much longer than the decay length of the bulk states, $\Delta S_E$ converges. On the other hand, in the case of $L_y=L_x$, the four corner states form a quadripartite entangled state which make additional contributions to $\Delta S_E$. Interestingly, if we take the difference between the $L_y=L_x$ and $L_y\gg L_x$ cases for large enough lattice sizes, we obtain a universal value $0.2797$ which is independent of the specific model parameters and only manifests the long range quadripartite entanglement among these four corner states.

Next, we use a four-sites toy model to explain the numerically obtained universal value $0.2797$. By solving the tight binding model containing only four corner states denoted by 1, 2, 3 and 4, respectively as shown in Fig.~\ref{fig:quadripartite}(a), with only nearest neighbour hopping and a threaded flux (in order to break the degeneracy), \cite{Benalcazar_S_2017} its two occupied states are obtained 
\begin{eqnarray}
  |\Psi_1\rangle&=&\frac12\left(|1\rangle-|2\rangle+|3\rangle-|4\rangle\right) \nonumber\\
  |\Psi_2\rangle&=&\frac12\left(|1\rangle-i|2\rangle-|3\rangle+i|4\rangle\right) \nonumber
\end{eqnarray}
The total ground state is then given by
\begin{eqnarray}
  |\Psi\rangle=|\Psi_1\rangle \times |\Psi_2\rangle,
\end{eqnarray}
which is a fully quadripartite entangled state in space. \footnote{Of course, in Fork space, it is a direct product state as a result of the non interacting model we used.} Then the eigenvalues of the reduced correlation matrices can be solved exactly as
\begin{eqnarray}
  \lambda_{12}=\lambda_{14}=\left\{\frac{1}{2}+\frac{\sqrt{2}}{4},\frac{1}{2}-\frac{\sqrt{2}}{4}\right\} \nonumber
\end{eqnarray}
and
\begin{eqnarray}
  \lambda_{13}=\left\{\frac{1}{2},\frac{1}{2}\right\} \nonumber
\end{eqnarray}
which then give the quadripartite entanglement entropy
\begin{eqnarray}
  \Delta S_E= 4\ln2-\sqrt{2}\ln\frac{2+\sqrt{2}}{2-\sqrt{2}}=0.2797
\end{eqnarray}
The value is exactly the same as what we obtained in numerics by subtracting the result of $L_y\gg L_x\rightarrow\infty$ from the square case with $L_y=L_x\rightarrow\infty$. In fact in the toy model by setting $L_y\rightarrow\infty$, \ie turning off the hoppings between sites $14$ and
$23$, the ground state is simply a direct product state
\begin{eqnarray}
  |\Psi\rangle=\frac12 \left(|1\rangle+|4\rangle\right) \times \left(|2\rangle+|3\rangle\right)
\end{eqnarray}
which indeed gives $\Delta S_E=0$ by definition. 

\section{Summary}

In this work, we studied the entanglement behavior of a second order topological insulator on the square lattice. For the bipartite entanglement spectra, we find that suitable chosen of the subsystem shape, \ie right angle corners matching the lattice in our model, would give zero modes in the entanglement spectrum corresponding to the zero energy corner states. For a finite size system, the long range entanglement among these four corner states are also studied. We propose a quadripartite scheme to measure the multipartite entanglement entropy which is well described by a four-sites toy model and thus can in turn used to identify the existence of these zero energy corner states. 

At the end, let us briefly discuss the generalizations to other types of higher order topological insulators. (1)For the bipartite entanglement spectra, the generalization is straightforward as long as the entanglement boundary is chosen the same as the physical boundary following Fidkowski's work. \cite{Fidkowski_PRL_2010} (2) In the case of open systems, the multipartite strategies depend on given geometries. For example, on the cubic lattice, we can divide the whole lattice into eight blocks (as a $2\times2\times2$ magic cube) denoting as $A_{1,2}$, $B_{1,2}$, $C_{1,2}$ and $D_{1,2}$ where the subscripts $(1,2)$ indicate two layers and $A\sim D$ are denoted following Fig.~\ref{fig:quadripartite}(a). Then, the subleading correction of the entanglement area law can be defined as $\Delta S_E= S_E(A_1,B_1,C_1,D_1)+S_E(A_1,B_1,A_2,B_2)+S_E(A_1,D_1,A_2,D_2)-S_E(A_1,C_1,B_2,D_2)$ whose nonzero value after removing relevant corner contributions can be used to identify the eight zero energy corner states. Similarly, different divide schemes can be designed for other geometries. 

\section{acknowledgement}
This project is supported by NSFC under Grant Nos. 11504164 and 11574134.

{\it Note added.} After completing our work, we notice another interesting paper \cite{Fukui_PRB_2018} on the similar topic.

\bibliography{ee2ti}

\begin{thebibliography}{35}%
\makeatletter
\providecommand \@ifxundefined [1]{%
 \@ifx{#1\undefined}
}%
\providecommand \@ifnum [1]{%
 \ifnum #1\expandafter \@firstoftwo
 \else \expandafter \@secondoftwo
 \fi
}%
\providecommand \@ifx [1]{%
 \ifx #1\expandafter \@firstoftwo
 \else \expandafter \@secondoftwo
 \fi
}%
\providecommand \natexlab [1]{#1}%
\providecommand \enquote  [1]{``#1''}%
\providecommand \bibnamefont  [1]{#1}%
\providecommand \bibfnamefont [1]{#1}%
\providecommand \citenamefont [1]{#1}%
\providecommand \href@noop [0]{\@secondoftwo}%
\providecommand \href [0]{\begingroup \@sanitize@url \@href}%
\providecommand \@href[1]{\@@startlink{#1}\@@href}%
\providecommand \@@href[1]{\endgroup#1\@@endlink}%
\providecommand \@sanitize@url [0]{\catcode `\\12\catcode `\$12\catcode
  `\&12\catcode `\#12\catcode `\^12\catcode `\_12\catcode `\%12\relax}%
\providecommand \@@startlink[1]{}%
\providecommand \@@endlink[0]{}%
\providecommand \url  [0]{\begingroup\@sanitize@url \@url }%
\providecommand \@url [1]{\endgroup\@href {#1}{\urlprefix }}%
\providecommand \urlprefix  [0]{URL }%
\providecommand \Eprint [0]{\href }%
\providecommand \doibase [0]{http://dx.doi.org/}%
\providecommand \selectlanguage [0]{\@gobble}%
\providecommand \bibinfo  [0]{\@secondoftwo}%
\providecommand \bibfield  [0]{\@secondoftwo}%
\providecommand \translation [1]{[#1]}%
\providecommand \BibitemOpen [0]{}%
\providecommand \bibitemStop [0]{}%
\providecommand \bibitemNoStop [0]{.\EOS\space}%
\providecommand \EOS [0]{\spacefactor3000\relax}%
\providecommand \BibitemShut  [1]{\csname bibitem#1\endcsname}%
\let\auto@bib@innerbib\@empty
\bibitem [{\citenamefont {Kitaev}\ and\ \citenamefont
  {Preskill}(2006)}]{Kitaev_PRL_2006}%
  \BibitemOpen
  \bibfield  {author} {\bibinfo {author} {\bibfnamefont {A.}~\bibnamefont
  {Kitaev}}\ and\ \bibinfo {author} {\bibfnamefont {J.}~\bibnamefont
  {Preskill}},\ }\href {\doibase 10.1103/PhysRevLett.96.110404} {\bibfield
  {journal} {\bibinfo  {journal} {Phys. Rev. Lett.}\ }\textbf {\bibinfo
  {volume} {96}},\ \bibinfo {pages} {110404} (\bibinfo {year}
  {2006})}\BibitemShut {NoStop}%
\bibitem [{\citenamefont {Levin}\ and\ \citenamefont
  {Wen}(2006)}]{Levin_PRL_2006}%
  \BibitemOpen
  \bibfield  {author} {\bibinfo {author} {\bibfnamefont {M.}~\bibnamefont
  {Levin}}\ and\ \bibinfo {author} {\bibfnamefont {X.-G.}\ \bibnamefont
  {Wen}},\ }\href {\doibase 10.1103/PhysRevLett.96.110405} {\bibfield
  {journal} {\bibinfo  {journal} {Phys. Rev. Lett.}\ }\textbf {\bibinfo
  {volume} {96}},\ \bibinfo {pages} {110405} (\bibinfo {year}
  {2006})}\BibitemShut {NoStop}%
\bibitem [{\citenamefont {Ryu}\ and\ \citenamefont
  {Hatsugai}(2006)}]{Ryu_PRB_2006}%
  \BibitemOpen
  \bibfield  {author} {\bibinfo {author} {\bibfnamefont {S.}~\bibnamefont
  {Ryu}}\ and\ \bibinfo {author} {\bibfnamefont {Y.}~\bibnamefont {Hatsugai}},\
  }\href {\doibase 10.1103/PhysRevB.73.245115} {\bibfield  {journal} {\bibinfo
  {journal} {Phys. Rev. B}\ }\textbf {\bibinfo {volume} {73}},\ \bibinfo
  {pages} {245115} (\bibinfo {year} {2006})}\BibitemShut {NoStop}%
\bibitem [{\citenamefont {Li}\ and\ \citenamefont
  {Haldane}(2008)}]{Li_PRL_2008}%
  \BibitemOpen
  \bibfield  {author} {\bibinfo {author} {\bibfnamefont {H.}~\bibnamefont
  {Li}}\ and\ \bibinfo {author} {\bibfnamefont {F.~D.~M.}\ \bibnamefont
  {Haldane}},\ }\href {\doibase 10.1103/PhysRevLett.101.010504} {\bibfield
  {journal} {\bibinfo  {journal} {Phys. Rev. Lett.}\ }\textbf {\bibinfo
  {volume} {101}},\ \bibinfo {pages} {010504} (\bibinfo {year}
  {2008})}\BibitemShut {NoStop}%
\bibitem [{\citenamefont {Fidkowski}(2010)}]{Fidkowski_PRL_2010}%
  \BibitemOpen
  \bibfield  {author} {\bibinfo {author} {\bibfnamefont {L.}~\bibnamefont
  {Fidkowski}},\ }\href {\doibase 10.1103/PhysRevLett.104.130502} {\bibfield
  {journal} {\bibinfo  {journal} {Phys. Rev. Lett.}\ }\textbf {\bibinfo
  {volume} {104}},\ \bibinfo {pages} {130502} (\bibinfo {year}
  {2010})}\BibitemShut {NoStop}%
\bibitem [{\citenamefont {Pollmann}\ \emph {et~al.}(2010)\citenamefont
  {Pollmann}, \citenamefont {Turner}, \citenamefont {Berg},\ and\ \citenamefont
  {Oshikawa}}]{Pollmann_PRB_2010}%
  \BibitemOpen
  \bibfield  {author} {\bibinfo {author} {\bibfnamefont {F.}~\bibnamefont
  {Pollmann}}, \bibinfo {author} {\bibfnamefont {A.~M.}\ \bibnamefont
  {Turner}}, \bibinfo {author} {\bibfnamefont {E.}~\bibnamefont {Berg}}, \ and\
  \bibinfo {author} {\bibfnamefont {M.}~\bibnamefont {Oshikawa}},\ }\href
  {\doibase 10.1103/PhysRevB.81.064439} {\bibfield  {journal} {\bibinfo
  {journal} {Phys. Rev. B}\ }\textbf {\bibinfo {volume} {81}},\ \bibinfo
  {pages} {064439} (\bibinfo {year} {2010})}\BibitemShut {NoStop}%
\bibitem [{\citenamefont {Yao}\ and\ \citenamefont {Qi}(2010)}]{Yao_PRL_2010}%
  \BibitemOpen
  \bibfield  {author} {\bibinfo {author} {\bibfnamefont {H.}~\bibnamefont
  {Yao}}\ and\ \bibinfo {author} {\bibfnamefont {X.-L.}\ \bibnamefont {Qi}},\
  }\href {\doibase 10.1103/PhysRevLett.105.080501} {\bibfield  {journal}
  {\bibinfo  {journal} {Phys. Rev. Lett.}\ }\textbf {\bibinfo {volume} {105}},\
  \bibinfo {pages} {080501} (\bibinfo {year} {2010})}\BibitemShut {NoStop}%
\bibitem [{\citenamefont {Turner}\ \emph {et~al.}(2011)\citenamefont {Turner},
  \citenamefont {Pollmann},\ and\ \citenamefont {Berg}}]{Turner_PRB_2011}%
  \BibitemOpen
  \bibfield  {author} {\bibinfo {author} {\bibfnamefont {A.~M.}\ \bibnamefont
  {Turner}}, \bibinfo {author} {\bibfnamefont {F.}~\bibnamefont {Pollmann}}, \
  and\ \bibinfo {author} {\bibfnamefont {E.}~\bibnamefont {Berg}},\ }\href
  {\doibase 10.1103/PhysRevB.83.075102} {\bibfield  {journal} {\bibinfo
  {journal} {Phys. Rev. B}\ }\textbf {\bibinfo {volume} {83}},\ \bibinfo
  {pages} {075102} (\bibinfo {year} {2011})}\BibitemShut {NoStop}%
\bibitem [{\citenamefont {Zhang}\ \emph {et~al.}(2011)\citenamefont {Zhang},
  \citenamefont {Grover},\ and\ \citenamefont {Vishwanath}}]{Zhang_PRL_2011}%
  \BibitemOpen
  \bibfield  {author} {\bibinfo {author} {\bibfnamefont {Y.}~\bibnamefont
  {Zhang}}, \bibinfo {author} {\bibfnamefont {T.}~\bibnamefont {Grover}}, \
  and\ \bibinfo {author} {\bibfnamefont {A.}~\bibnamefont {Vishwanath}},\
  }\href {\doibase 10.1103/PhysRevLett.107.067202} {\bibfield  {journal}
  {\bibinfo  {journal} {Phys. Rev. Lett.}\ }\textbf {\bibinfo {volume} {107}},\
  \bibinfo {pages} {067202} (\bibinfo {year} {2011})}\BibitemShut {NoStop}%
\bibitem [{\citenamefont {Huang}\ and\ \citenamefont
  {Arovas}(2012)}]{Huang_PRB_2012}%
  \BibitemOpen
  \bibfield  {author} {\bibinfo {author} {\bibfnamefont {Z.}~\bibnamefont
  {Huang}}\ and\ \bibinfo {author} {\bibfnamefont {D.~P.}\ \bibnamefont
  {Arovas}},\ }\href {\doibase 10.1103/PhysRevB.86.245109} {\bibfield
  {journal} {\bibinfo  {journal} {Phys. Rev. B}\ }\textbf {\bibinfo {volume}
  {86}},\ \bibinfo {pages} {245109} (\bibinfo {year} {2012})}\BibitemShut
  {NoStop}%
\bibitem [{\citenamefont {Qi}\ \emph {et~al.}(2012)\citenamefont {Qi},
  \citenamefont {Katsura},\ and\ \citenamefont {Ludwig}}]{Qi_PRL_2012}%
  \BibitemOpen
  \bibfield  {author} {\bibinfo {author} {\bibfnamefont {X.-L.}\ \bibnamefont
  {Qi}}, \bibinfo {author} {\bibfnamefont {H.}~\bibnamefont {Katsura}}, \ and\
  \bibinfo {author} {\bibfnamefont {A.~W.~W.}\ \bibnamefont {Ludwig}},\ }\href
  {\doibase 10.1103/PhysRevLett.108.196402} {\bibfield  {journal} {\bibinfo
  {journal} {Phys. Rev. Lett.}\ }\textbf {\bibinfo {volume} {108}},\ \bibinfo
  {pages} {196402} (\bibinfo {year} {2012})}\BibitemShut {NoStop}%
\bibitem [{\citenamefont {Jiang}\ \emph {et~al.}(2013)\citenamefont {Jiang},
  \citenamefont {Singh},\ and\ \citenamefont {Balents}}]{Jiang_PRL_2013}%
  \BibitemOpen
  \bibfield  {author} {\bibinfo {author} {\bibfnamefont {H.-C.}\ \bibnamefont
  {Jiang}}, \bibinfo {author} {\bibfnamefont {R.~R.~P.}\ \bibnamefont {Singh}},
  \ and\ \bibinfo {author} {\bibfnamefont {L.}~\bibnamefont {Balents}},\ }\href
  {\doibase 10.1103/PhysRevLett.111.107205} {\bibfield  {journal} {\bibinfo
  {journal} {Phys. Rev. Lett.}\ }\textbf {\bibinfo {volume} {111}},\ \bibinfo
  {pages} {107205} (\bibinfo {year} {2013})}\BibitemShut {NoStop}%
\bibitem [{\citenamefont {Chandran}\ \emph {et~al.}(2014)\citenamefont
  {Chandran}, \citenamefont {Khemani},\ and\ \citenamefont
  {Sondhi}}]{Chandran_PRL_2014}%
  \BibitemOpen
  \bibfield  {author} {\bibinfo {author} {\bibfnamefont {A.}~\bibnamefont
  {Chandran}}, \bibinfo {author} {\bibfnamefont {V.}~\bibnamefont {Khemani}}, \
  and\ \bibinfo {author} {\bibfnamefont {S.~L.}\ \bibnamefont {Sondhi}},\
  }\href {\doibase 10.1103/PhysRevLett.113.060501} {\bibfield  {journal}
  {\bibinfo  {journal} {Phys. Rev. Lett.}\ }\textbf {\bibinfo {volume} {113}},\
  \bibinfo {pages} {060501} (\bibinfo {year} {2014})}\BibitemShut {NoStop}%
\bibitem [{\citenamefont {Wang}\ \emph {et~al.}(2015)\citenamefont {Wang},
  \citenamefont {Xu}, \citenamefont {Wang},\ and\ \citenamefont
  {Wu}}]{Wang_PRB_2015}%
  \BibitemOpen
  \bibfield  {author} {\bibinfo {author} {\bibfnamefont {D.}~\bibnamefont
  {Wang}}, \bibinfo {author} {\bibfnamefont {S.}~\bibnamefont {Xu}}, \bibinfo
  {author} {\bibfnamefont {Y.}~\bibnamefont {Wang}}, \ and\ \bibinfo {author}
  {\bibfnamefont {C.}~\bibnamefont {Wu}},\ }\href {\doibase
  10.1103/PhysRevB.91.115118} {\bibfield  {journal} {\bibinfo  {journal} {Phys.
  Rev. B}\ }\textbf {\bibinfo {volume} {91}},\ \bibinfo {pages} {115118}
  (\bibinfo {year} {2015})}\BibitemShut {NoStop}%
\bibitem [{\citenamefont {Zeng}\ \emph {et~al.}(2015)\citenamefont {Zeng},
  \citenamefont {Chen}, \citenamefont {Zhou},\ and\ \citenamefont
  {Wen}}]{Zeng_a_2015}%
  \BibitemOpen
  \bibfield  {author} {\bibinfo {author} {\bibfnamefont {B.}~\bibnamefont
  {Zeng}}, \bibinfo {author} {\bibfnamefont {X.}~\bibnamefont {Chen}}, \bibinfo
  {author} {\bibfnamefont {D.-L.}\ \bibnamefont {Zhou}}, \ and\ \bibinfo
  {author} {\bibfnamefont {X.-G.}\ \bibnamefont {Wen}},\ }\href@noop {}
  {\bibfield  {journal} {\bibinfo  {journal} {arXiv:1508.02595}\ } (\bibinfo
  {year} {2015})},\ \Eprint {http://arxiv.org/abs/1508.02595}
  {arXiv:1508.02595} \BibitemShut {NoStop}%
\bibitem [{\citenamefont {Laflorencie}(2016)}]{Laflorencie_PR_2016}%
  \BibitemOpen
  \bibfield  {author} {\bibinfo {author} {\bibfnamefont {N.}~\bibnamefont
  {Laflorencie}},\ }\href {\doibase 10.1016/j.physrep.2016.06.008} {\bibfield
  {journal} {\bibinfo  {journal} {Phys. Rep.}\ }\textbf {\bibinfo {volume}
  {646}},\ \bibinfo {pages} {1} (\bibinfo {year} {2016})}\BibitemShut {NoStop}%
\bibitem [{\citenamefont {{Koch-Janusz}}\ \emph {et~al.}(2017)\citenamefont
  {{Koch-Janusz}}, \citenamefont {Dhochak},\ and\ \citenamefont
  {Berg}}]{Koch-Janusz_PRB_2017}%
  \BibitemOpen
  \bibfield  {author} {\bibinfo {author} {\bibfnamefont {M.}~\bibnamefont
  {{Koch-Janusz}}}, \bibinfo {author} {\bibfnamefont {K.}~\bibnamefont
  {Dhochak}}, \ and\ \bibinfo {author} {\bibfnamefont {E.}~\bibnamefont
  {Berg}},\ }\href {\doibase 10.1103/PhysRevB.95.205110} {\bibfield  {journal}
  {\bibinfo  {journal} {Phys. Rev. B}\ }\textbf {\bibinfo {volume} {95}},\
  \bibinfo {pages} {205110} (\bibinfo {year} {2017})}\BibitemShut {NoStop}%
\bibitem [{\citenamefont {Hatsugai}(1993)}]{Hatsugai_PRL_1993}%
  \BibitemOpen
  \bibfield  {author} {\bibinfo {author} {\bibfnamefont {Y.}~\bibnamefont
  {Hatsugai}},\ }\href {\doibase 10.1103/PhysRevLett.71.3697} {\bibfield
  {journal} {\bibinfo  {journal} {Phys. Rev. Lett.}\ }\textbf {\bibinfo
  {volume} {71}},\ \bibinfo {pages} {3697} (\bibinfo {year}
  {1993})}\BibitemShut {NoStop}%
\bibitem [{\citenamefont {Qi}\ \emph {et~al.}(2006)\citenamefont {Qi},
  \citenamefont {Wu},\ and\ \citenamefont {Zhang}}]{Qi_PRB_2006}%
  \BibitemOpen
  \bibfield  {author} {\bibinfo {author} {\bibfnamefont {X.-L.}\ \bibnamefont
  {Qi}}, \bibinfo {author} {\bibfnamefont {Y.-S.}\ \bibnamefont {Wu}}, \ and\
  \bibinfo {author} {\bibfnamefont {S.-C.}\ \bibnamefont {Zhang}},\ }\href
  {\doibase 10.1103/PhysRevB.74.045125} {\bibfield  {journal} {\bibinfo
  {journal} {Phys. Rev. B}\ }\textbf {\bibinfo {volume} {74}},\ \bibinfo
  {pages} {045125} (\bibinfo {year} {2006})}\BibitemShut {NoStop}%
\bibitem [{\citenamefont {Qi}\ \emph {et~al.}(2008)\citenamefont {Qi},
  \citenamefont {Hughes},\ and\ \citenamefont {Zhang}}]{Qi_PRB_2008}%
  \BibitemOpen
  \bibfield  {author} {\bibinfo {author} {\bibfnamefont {X.-L.}\ \bibnamefont
  {Qi}}, \bibinfo {author} {\bibfnamefont {T.~L.}\ \bibnamefont {Hughes}}, \
  and\ \bibinfo {author} {\bibfnamefont {S.-C.}\ \bibnamefont {Zhang}},\ }\href
  {\doibase 10.1103/PhysRevB.78.195424} {\bibfield  {journal} {\bibinfo
  {journal} {Phys. Rev. B}\ }\textbf {\bibinfo {volume} {78}},\ \bibinfo
  {pages} {195424} (\bibinfo {year} {2008})}\BibitemShut {NoStop}%
\bibitem [{\citenamefont {Schnyder}\ \emph {et~al.}(2008)\citenamefont
  {Schnyder}, \citenamefont {Ryu}, \citenamefont {Furusaki},\ and\
  \citenamefont {Ludwig}}]{Schnyder_PRB_2008}%
  \BibitemOpen
  \bibfield  {author} {\bibinfo {author} {\bibfnamefont {A.}~\bibnamefont
  {Schnyder}}, \bibinfo {author} {\bibfnamefont {S.}~\bibnamefont {Ryu}},
  \bibinfo {author} {\bibfnamefont {A.}~\bibnamefont {Furusaki}}, \ and\
  \bibinfo {author} {\bibfnamefont {A.}~\bibnamefont {Ludwig}},\ }\href
  {\doibase 10.1103/PhysRevB.78.195125} {\bibfield  {journal} {\bibinfo
  {journal} {Phys. Rev. B}\ }\textbf {\bibinfo {volume} {78}},\ \bibinfo
  {pages} {195125} (\bibinfo {year} {2008})}\BibitemShut {NoStop}%
\bibitem [{\citenamefont {Benalcazar}\ \emph
  {et~al.}(2017{\natexlab{a}})\citenamefont {Benalcazar}, \citenamefont
  {Bernevig},\ and\ \citenamefont {Hughes}}]{Benalcazar_S_2017}%
  \BibitemOpen
  \bibfield  {author} {\bibinfo {author} {\bibfnamefont {W.~A.}\ \bibnamefont
  {Benalcazar}}, \bibinfo {author} {\bibfnamefont {B.~A.}\ \bibnamefont
  {Bernevig}}, \ and\ \bibinfo {author} {\bibfnamefont {T.~L.}\ \bibnamefont
  {Hughes}},\ }\href {\doibase 10.1126/science.aah6442} {\bibfield  {journal}
  {\bibinfo  {journal} {Science}\ }\textbf {\bibinfo {volume} {357}},\ \bibinfo
  {pages} {61} (\bibinfo {year} {2017}{\natexlab{a}})}\BibitemShut {NoStop}%
\bibitem [{\citenamefont {Benalcazar}\ \emph
  {et~al.}(2017{\natexlab{b}})\citenamefont {Benalcazar}, \citenamefont
  {Bernevig},\ and\ \citenamefont {Hughes}}]{Benalcazar_PRB_2017}%
  \BibitemOpen
  \bibfield  {author} {\bibinfo {author} {\bibfnamefont {W.~A.}\ \bibnamefont
  {Benalcazar}}, \bibinfo {author} {\bibfnamefont {B.~A.}\ \bibnamefont
  {Bernevig}}, \ and\ \bibinfo {author} {\bibfnamefont {T.~L.}\ \bibnamefont
  {Hughes}},\ }\href {\doibase 10.1103/PhysRevB.96.245115} {\bibfield
  {journal} {\bibinfo  {journal} {Phys. Rev. B}\ }\textbf {\bibinfo {volume}
  {96}},\ \bibinfo {pages} {245115} (\bibinfo {year}
  {2017}{\natexlab{b}})}\BibitemShut {NoStop}%
\bibitem [{\citenamefont {Schindler}\ \emph
  {et~al.}(2018{\natexlab{a}})\citenamefont {Schindler}, \citenamefont {Cook},
  \citenamefont {Vergniory}, \citenamefont {Wang}, \citenamefont {Parkin},
  \citenamefont {Bernevig},\ and\ \citenamefont {Neupert}}]{Schindler_SA_2018}%
  \BibitemOpen
  \bibfield  {author} {\bibinfo {author} {\bibfnamefont {F.}~\bibnamefont
  {Schindler}}, \bibinfo {author} {\bibfnamefont {A.~M.}\ \bibnamefont {Cook}},
  \bibinfo {author} {\bibfnamefont {M.~G.}\ \bibnamefont {Vergniory}}, \bibinfo
  {author} {\bibfnamefont {Z.}~\bibnamefont {Wang}}, \bibinfo {author}
  {\bibfnamefont {S.~S.~P.}\ \bibnamefont {Parkin}}, \bibinfo {author}
  {\bibfnamefont {B.~A.}\ \bibnamefont {Bernevig}}, \ and\ \bibinfo {author}
  {\bibfnamefont {T.}~\bibnamefont {Neupert}},\ }\href {\doibase
  10.1126/sciadv.aat0346} {\bibfield  {journal} {\bibinfo  {journal} {Sci.
  Adv.}\ }\textbf {\bibinfo {volume} {4}},\ \bibinfo {pages} {eaat0346}
  (\bibinfo {year} {2018}{\natexlab{a}})}\BibitemShut {NoStop}%
\bibitem [{\citenamefont {Song}\ \emph {et~al.}(2017)\citenamefont {Song},
  \citenamefont {Fang},\ and\ \citenamefont {Fang}}]{Song_PRL_2017}%
  \BibitemOpen
  \bibfield  {author} {\bibinfo {author} {\bibfnamefont {Z.}~\bibnamefont
  {Song}}, \bibinfo {author} {\bibfnamefont {Z.}~\bibnamefont {Fang}}, \ and\
  \bibinfo {author} {\bibfnamefont {C.}~\bibnamefont {Fang}},\ }\href {\doibase
  10.1103/PhysRevLett.119.246402} {\bibfield  {journal} {\bibinfo  {journal}
  {Phys. Rev. Lett.}\ }\textbf {\bibinfo {volume} {119}},\ \bibinfo {pages}
  {246402} (\bibinfo {year} {2017})}\BibitemShut {NoStop}%
\bibitem [{\citenamefont {Parameswaran}\ and\ \citenamefont
  {Wan}(2017)}]{Parameswaran_P_2017}%
  \BibitemOpen
  \bibfield  {author} {\bibinfo {author} {\bibfnamefont {S.~A.}\ \bibnamefont
  {Parameswaran}}\ and\ \bibinfo {author} {\bibfnamefont {Y.}~\bibnamefont
  {Wan}},\ }\href@noop {} {\bibfield  {journal} {\bibinfo  {journal} {Physics}\
  }\textbf {\bibinfo {volume} {10}},\ \bibinfo {pages} {132} (\bibinfo {year}
  {2017})}\BibitemShut {NoStop}%
\bibitem [{\citenamefont {Xu}\ \emph {et~al.}(2017)\citenamefont {Xu},
  \citenamefont {Xue},\ and\ \citenamefont {Wan}}]{Xu_a_2017}%
  \BibitemOpen
  \bibfield  {author} {\bibinfo {author} {\bibfnamefont {Y.}~\bibnamefont
  {Xu}}, \bibinfo {author} {\bibfnamefont {R.}~\bibnamefont {Xue}}, \ and\
  \bibinfo {author} {\bibfnamefont {S.}~\bibnamefont {Wan}},\ }\href@noop {}
  {\bibfield  {journal} {\bibinfo  {journal} {arXiv:1711.09202}\ } (\bibinfo
  {year} {2017})},\ \Eprint {http://arxiv.org/abs/1711.09202}
  {arXiv:1711.09202} \BibitemShut {NoStop}%
\bibitem [{\citenamefont {Ezawa}(2018{\natexlab{a}})}]{Ezawa_PRL_2018}%
  \BibitemOpen
  \bibfield  {author} {\bibinfo {author} {\bibfnamefont {M.}~\bibnamefont
  {Ezawa}},\ }\href {\doibase 10.1103/PhysRevLett.120.026801} {\bibfield
  {journal} {\bibinfo  {journal} {Phys. Rev. Lett.}\ }\textbf {\bibinfo
  {volume} {120}},\ \bibinfo {pages} {026801} (\bibinfo {year}
  {2018}{\natexlab{a}})}\BibitemShut {NoStop}%
\bibitem [{\citenamefont {Ezawa}(2018{\natexlab{b}})}]{Ezawa_PRB_2018}%
  \BibitemOpen
  \bibfield  {author} {\bibinfo {author} {\bibfnamefont {M.}~\bibnamefont
  {Ezawa}},\ }\href {\doibase 10.1103/PhysRevB.97.155305} {\bibfield  {journal}
  {\bibinfo  {journal} {Phys. Rev. B}\ }\textbf {\bibinfo {volume} {97}},\
  \bibinfo {pages} {155305} (\bibinfo {year} {2018}{\natexlab{b}})}\BibitemShut
  {NoStop}%
\bibitem [{\citenamefont {Kunst}\ \emph {et~al.}(2018)\citenamefont {Kunst},
  \citenamefont {{van Miert}},\ and\ \citenamefont
  {Bergholtz}}]{Kunst_PRB_2018}%
  \BibitemOpen
  \bibfield  {author} {\bibinfo {author} {\bibfnamefont {F.~K.}\ \bibnamefont
  {Kunst}}, \bibinfo {author} {\bibfnamefont {G.}~\bibnamefont {{van Miert}}},
  \ and\ \bibinfo {author} {\bibfnamefont {E.~J.}\ \bibnamefont {Bergholtz}},\
  }\href {\doibase 10.1103/PhysRevB.97.241405} {\bibfield  {journal} {\bibinfo
  {journal} {Phys. Rev. B}\ }\textbf {\bibinfo {volume} {97}},\ \bibinfo
  {pages} {241405} (\bibinfo {year} {2018})}\BibitemShut {NoStop}%
\bibitem [{\citenamefont {Schindler}\ \emph
  {et~al.}(2018{\natexlab{b}})\citenamefont {Schindler}, \citenamefont {Wang},
  \citenamefont {Vergniory}, \citenamefont {Cook}, \citenamefont {Murani},
  \citenamefont {Sengupta}, \citenamefont {Kasumov}, \citenamefont {Deblock},
  \citenamefont {Jeon}, \citenamefont {Drozdov}, \citenamefont {Bouchiat},
  \citenamefont {Gu\'eron}, \citenamefont {Yazdani}, \citenamefont {Bernevig},\
  and\ \citenamefont {Neupert}}]{Schindler_a_2018}%
  \BibitemOpen
  \bibfield  {author} {\bibinfo {author} {\bibfnamefont {F.}~\bibnamefont
  {Schindler}}, \bibinfo {author} {\bibfnamefont {Z.}~\bibnamefont {Wang}},
  \bibinfo {author} {\bibfnamefont {M.~G.}\ \bibnamefont {Vergniory}}, \bibinfo
  {author} {\bibfnamefont {A.~M.}\ \bibnamefont {Cook}}, \bibinfo {author}
  {\bibfnamefont {A.}~\bibnamefont {Murani}}, \bibinfo {author} {\bibfnamefont
  {S.}~\bibnamefont {Sengupta}}, \bibinfo {author} {\bibfnamefont {A.~Y.}\
  \bibnamefont {Kasumov}}, \bibinfo {author} {\bibfnamefont {R.}~\bibnamefont
  {Deblock}}, \bibinfo {author} {\bibfnamefont {S.}~\bibnamefont {Jeon}},
  \bibinfo {author} {\bibfnamefont {I.}~\bibnamefont {Drozdov}}, \bibinfo
  {author} {\bibfnamefont {H.}~\bibnamefont {Bouchiat}}, \bibinfo {author}
  {\bibfnamefont {S.}~\bibnamefont {Gu\'eron}}, \bibinfo {author}
  {\bibfnamefont {A.}~\bibnamefont {Yazdani}}, \bibinfo {author} {\bibfnamefont
  {B.~A.}\ \bibnamefont {Bernevig}}, \ and\ \bibinfo {author} {\bibfnamefont
  {T.}~\bibnamefont {Neupert}},\ }\href@noop {} {\bibfield  {journal} {\bibinfo
   {journal} {arXiv:1802.02585}\ } (\bibinfo {year} {2018}{\natexlab{b}})},\
  \Eprint {http://arxiv.org/abs/1802.02585} {arXiv:1802.02585} \BibitemShut
  {NoStop}%
\bibitem [{\citenamefont {Peschel}(2003)}]{Peschel_JPAMG_2003}%
  \BibitemOpen
  \bibfield  {author} {\bibinfo {author} {\bibfnamefont {I.}~\bibnamefont
  {Peschel}},\ }\href {\doibase 10.1088/0305-4470/36/14/101} {\bibfield
  {journal} {\bibinfo  {journal} {J. Phys. A: Math. Gen.}\ }\textbf {\bibinfo
  {volume} {36}},\ \bibinfo {pages} {L205} (\bibinfo {year}
  {2003})}\BibitemShut {NoStop}%
\bibitem [{Note1()}]{Note1}%
  \BibitemOpen
  \bibinfo {note} {Exactly speaking, all physical systems are finite and the
  edge or corner states will be entangled together by long range
  hybridizations. However, on the other hand, these physical systems are also
  affected by some environment, which may cause quantum decoherence and thus
  break down these long range entangled states. Therefore, here, when we say a
  finite size system, we just mean that the size is smaller than the
  decoherence length.}\BibitemShut {Stop}%
\bibitem [{Note2()}]{Note2}%
  \BibitemOpen
  \bibinfo {note} {Of course, in Fork space, it is a direct product state as a
  result of the non interacting model we used.}\BibitemShut {Stop}%
\bibitem [{\citenamefont {Fukui}\ and\ \citenamefont
  {Hatsugai}(2018)}]{Fukui_PRB_2018}%
  \BibitemOpen
  \bibfield  {author} {\bibinfo {author} {\bibfnamefont {T.}~\bibnamefont
  {Fukui}}\ and\ \bibinfo {author} {\bibfnamefont {Y.}~\bibnamefont
  {Hatsugai}},\ }\href {\doibase 10.1103/PhysRevB.98.035147} {\bibfield
  {journal} {\bibinfo  {journal} {Phys. Rev. B}\ }\textbf {\bibinfo {volume}
  {98}},\ \bibinfo {pages} {035147} (\bibinfo {year} {2018})}\BibitemShut
  {NoStop}%
\end{thebibliography}%
\end{document}